\documentclass[aps,prl,floatfix,10pt]{revtex4-1}
\usepackage[utf8]{inputenc}
\usepackage{mathrsfs}
\usepackage{amsmath}
\usepackage{amsfonts}
\usepackage{amssymb}
\usepackage{amsthm}
\usepackage{multirow}
\usepackage{color}
\usepackage{graphicx}
\usepackage{lettrine}
\graphicspath{{./images/}}
\usepackage{geometry}
\usepackage{hyperref}
\usepackage{lineno}
\usepackage{textcomp}
\usepackage{bbding}
\usepackage{pifont}

\hypersetup{
     unicode=false,          
     pdftoolbar=true,        
     pdfmenubar=true,        
     pdffitwindow=false,     
     pdfstartview={FitH},    
     pdftitle={My title},    
     pdfauthor={Author},     
     pdfsubject={Subject},   
     pdfcreator={Creator},   
     pdfproducer={Producer}, 
     pdfkeywords={keyword1} {key2} {key3}, 
     pdfnewwindow=true,      
     colorlinks=false,       
     linkcolor=red,
     citecolor=green,        
     filecolor=magenta,      
     urlcolor=cyan           
}
\usepackage[toc,page]{appendix}
\usepackage{ulem}

\begin{document}

\onecolumngrid
\begin{flushleft}
{\Large{Stripe skyrmions and skyrmion crystals}}

\quad\par
\quad\par
X. R. Wang$^{1,2,*}$, X. C. Hu$^{1,2}$, and H. T. Wu$^{1,2}$
\quad\par
\quad\par
\begin{enumerate}
\item Physics Department, The Hong Kong University of Science and Technology,
Clear Water Bay, Kowloon, Hong Kong.
\item HKUST Shenzhen Research Institute, Shenzhen 518057, China.
\end{enumerate}
\noindent{Correspondence and requests for materials should be addressed
to X.R.W. (email: phxwan@ust.hk)}
\end{flushleft}
\quad\par
\quad\par
{\bf
Skyrmions are important in topological quantum field theory for being soliton solutions 
of a nonlinear sigma model and in information technology for their attractive applications. 
Skyrmions are believed to be circular and stripy spin textures appeared in the vicinity 
of skyrmion crystals are termed spiral, helical, and cycloid spin orders, but not skyrmions. 
Here we present convincing evidences showing that those stripy spin textures are skyrmions, 
``siblings" of circular skyrmions in skyrmion crystals and ``cousins" of isolated circular skyrmions. Specifically, isolated skyrmions are excitations when skyrmion formation energy 
is positive. The skyrmion morphologies are various stripy structures when the ground 
states of chiral magnetic films are skyrmions. 
The density of skyrmion number determines the morphology of condensed skyrmion states. 
At the extreme of one skyrmion in the whole sample, the skyrmion is a ramified stripe. 
As the skyrmion number density increases, individual skyrmion shapes gradually change from 
ramified stripes to rectangular stripes, and eventually to disk-like objects. 
At a low skyrmion number density, the natural width of stripes is proportional to the ratio 
between the exchange stiffness constant and Dzyaloshinskii-Moriya interaction coefficient. 
At a high skyrmion number density, skyrmion crystals are the preferred states. 
Our findings reveal the nature and properties of stripy spin texture, and open a new avenue 
for manipulating skyrmions, especially condensed skyrmions such as skyrmion crystals.} 
\newpage
\twocolumngrid
Skyrmions, originally used to describe resonance states of baryons \cite{skyrme}, were 
first unambiguously observed in the form of skyrmion crystals in various chiral magnets and by various experimental techniques \cite{Muhlbauer2009,Yu2010,Yu2011,Heinze,Romming,Onose}. 
Magnetic skyrmions are topological non-trivial spin textures of magnetic films 
characterized by a skyrmion number of $Q=\frac{1}{4\pi}\int \vec{m}\cdot(\partial_x \vec{m}
\times \partial_y \vec{m}){\rm d}x{\rm d}y$, here $\vec{m}$ is the unit vector of magnetization. 
$Q$ must be an integer for an infinite magnetic film, and a non-zero $Q$ spin texture is 
called a skyrmion of skyrmion number $Q$. It was isolated circular skyrmions, not a 
skyrmion crystal, that were predicted in early theories \cite{Bogdanov2001,Rossler2006}.  
Isolated skyrmions were indeed observed later in confined structures and films 
\cite{Fert,Li,roadmap,Jiang,Tian,Yuan2016}. It is an experimental fact that 
skyrmion crystals form in very narrow magnetic-field-temperature windows. 
Outside of the windows, stripy phases appear. The stripy phases, which can even coexist 
with skyrmion crystals, are in fact easier to form than a skyrmion crystal does. 
In contrast, those stripy phases do not appear together with isolated skyrmions. 
Interestingly, the stripy phase was observed many years before the observations of 
skyrmion crystals \cite{Uchida}. These stripy phases are called helical, spiral, and 
cycloid spin orders.  A one-dimensional model \cite{Rossler2006, Han2010, Leonov2010, 
Thiaville2013, Romming2016, Bogdanov1994} was used to describe the rotation of spins 
perpendicular to stripes. 
To date, a holistic description of various stripy structures, especially ramified stripes and stripy maze, \cite{Jiang2015, 
Yu2018, Birch2020, He2017, Kumar2020, Huang2017, Raju2019, Schoenherr2018} is lacking.  
The general belief is that those stripes are not skyrmions and have zero skyrmion 
numbers \cite{Wiesendanger} although some race-track stripes are called 
merons with 1/2 skyrmion number or bimerons\cite{Ezawa2011, Silva2014, Huang2017}.

In this letter we show that stripy magnetic textures appear in a chiral magnetic 
film with Dzyaloshinskii-Moriya interaction (DMI) are actually irregular skyrmions. 
Each stripy texture has exactly skyrmion number 1. The irregular shape is due to the negative 
skyrmion formation energy (relative to ferromagnetic state) when the ferromagnetic state is 
not the ground state. For a given system, the morphology of these skyrmions are random when the 
skyrmion number density is low. At extremely low density, magnetic textures are ramified stripes. 
The exact appearance of each pattern is very sensitive to the initial spin configuration 
and its dynamical path. The basic building blocks of irregular random skyrmions are stripes 
of well-defined width. The optimal width comes from the competition between the Heisenberg's 
exchange energy and the DMI energy that respectively prefer a larger and smaller width in order 
to minimize exchange energy cost and maximize the negative skyrmion formation energy gain.
Unexpectedly, this exchange energy and DMI energy dependence of width is opposite to the skyrmion 
size of an isolated skyrmion that increases with DMI interaction and decreases with exchange energy \cite{Xiansi}. 

We consider an ultra-thin ferromagnetic film of thickness $d$ in the $xy$ plane. 
The film has an exchange energy $E_\mathrm{ex}$ with exchange stiffness constant $A$, an 
interfacial DMI energy $E_\mathrm{DM}$ with DMI coefficient $D$, an anisotropy energy 
$E_\mathrm{an}$ with a perpendicular easy-axis anisotropy $K$, and the Zeeman energy 
$E_\mathrm{Ze}$ in a perpendicular magnetic field $H$. The total energy $E_{total}$ reads 
\begin{equation}
E_{total}=E_\mathrm{ex}+E_\mathrm{DM}+E_\mathrm{an}+E_\mathrm{Ze},
\label{energy}
\end{equation}
where $E_\mathrm{ex}=Ad\iint |\nabla \vec{m}|^2\mathrm{d}S$, $E_\mathrm{DM}=Dd\iint 
[m_z\nabla\cdot\vec{m}-(\vec{m}\cdot\nabla)m_z]\mathrm{d}S$, $E_\mathrm{an}=Kd\iint 
(1-m_z^2)\mathrm{d}S$, and $E_\mathrm{Ze}=\mu_0HM_\text{s}d\iint (1-m_z)\mathrm{d}S$. 
$M_\text{s}$ is the saturation magnetization and $\mu_0$ is the vacuum permeability. 
The integration is over the whole film. The energy is set to zero, $E_{total}=0$, for the  
ferromagnetic state of $m_z=1$. We have assumed $\vec{m}$ is uniform in thickness direction. 
The demagnetization effect is included in $E_\mathrm{an}$ through the effective anisotropy 
$K=K_u-\mu_0M_\text{s}^2/2$ corrected by the shape anisotropy, here $K_u$ is the perpendicular 
magnetocrystalline anisotropy. This is a good approximation when the film thickness $d$ is 
much smaller than the exchange length \cite{Xiansi}. 
It is known that isolated circular skyrmions are metastable state of energy 
$8\pi Ad \sqrt{1-\kappa}$ when $\kappa=\pi^2 D^2/(16AK)<1$ \cite{Xiansi}. 
Here we use MuMax3 simulator \cite{MuMax3} to numerically solve the Landau-Lifshitz-Gilbert 
(LLG) equation for the stable states in the opposite regime of $\kappa > 1 $ where circular 
skyrmions are not stable states \cite{Xiansi}, see Method. 

\begin{figure*}
\centering
\includegraphics[width=15cm]{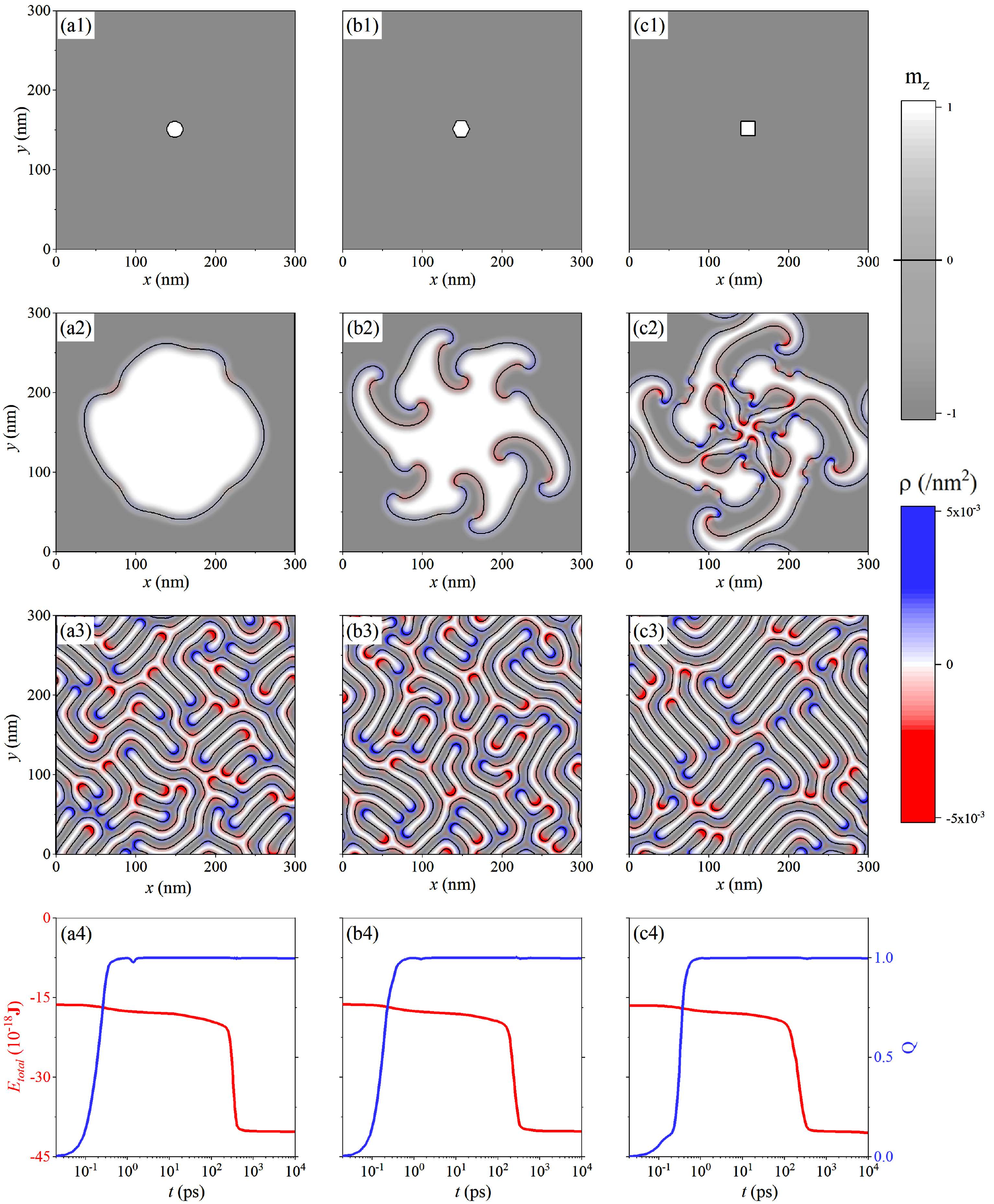}\\
\caption{Numerical solutions of LLG equation under the periodical boundary conditions 
and with $A=10{\rm pJ/m}, D=6{\rm mJ/m^2}, K=0.49{\rm MJ/m^3}, M_s=0.58{\rm MA/m}$ 
for a sample of $300\rm{nm}\times 300\rm{nm}\times 0.4\rm{nm}$. 
(a1,b1,c1) are different initial configurations of a disk of diameter $20 \rm{nm}$ (a1), a 
hexagon of side length $10 \rm{nm}$ (b1) and a square of length $20 \rm{nm}$ (c1). (a2,b2,c2) are 
intermediate states at 0.3ns with irregular shapes due to the negative formation energy.  
(a3,b3,c3) are the final stable pattern with irregular ramified stripes. 
Skyrmion charge density $\rho$ is encoded by colours (the blue for positive and the red for 
negative) while the gray-scale encodes $m_z$. The dark black lines denote $m_z=0$.
The positive and negative charges exist respectively only around convex and concave areas. 
The spin profile across the stripes at the green \textcircled{n} is considered. 
(a4, b4, c4) are the evolution of total energy $E_{total}$ and topological skyrmion number 
$Q$ ($t$ is in the logarithmic scale). $Q$ reaches the skyrmion number 1 within $1 ps$. 
Clearly, the skyrmion number is a constant and the total energy is negative 
and approaches a constant almost independent from the initial configurations.}
\label{fig1}
\end{figure*}

For a sample of $300\rm{nm}\times 300\rm{nm}\times 0.4\rm{nm}$ under the periodical boundary 
conditions and with $A=10{\rm pJ/m}, D=6\rm{mJ/m^2}, K=0.49\rm{MJ/m^3}$, 
and $M_s=0.58\rm{MA/m}$ that are typical values for those chiral magnets supporting 
skyrmion crystals \cite{Fert2,zhou-2020,Wiesendanger}, $\kappa > 1 $, so that 
single domain and isolated circular skyrmion are not stable any more \cite{Xiansi}.  
We will start from a small nucleation magnetic domain with sharp domain wall to speed up 
skyrmion formation dynamics although stripe skyrmions will appear spontaneously due to 
thermal or other fluctuations in reality. Figure \ref{fig1} shows how various initial states 
in the top panel (a1, b1, c1) evolve according to the LLG equation with $\alpha=0.25$. 
$\alpha$ will not change physics described below because we are interested in the spin 
textures of the LLG equation that do not vary with time. However, $\alpha$ can change the 
evolution path and energy dissipation rate so that it can change the intermediate states 
shown in (a2-c2) and influence which fixed point to pick when a system has many fixed points 
(textures) like the current situation. This is similar to the sensitiveness of attractor 
basins to the damping in a macro spin system \cite{zzsun}. The initial configurations 
are obtained by reversing spins in the white regions in (a1-c1) from $m_z=-1$ to $m_z=1$ 
such that the configurations have very sharp domain walls and zero skyrmion number $Q=0$. 
After a short time of an order of picosecond, the initial states transform into irregular 
structures of skyrmion number $Q=1$, no matter whether the initial shape is circular or 
non-circular as shown in (a4-c4) where the time is in logarithmic scale. As time goes on,  
$Q$ (the blue lines in a4-c4) stays at 1, and system energy $E_{total}$ (the red curves) 
is negative and keeps decreasing until it reaches a stable ramified stripy spin texture. 
Clearly, this irregular ramified spin texture is a non-circular skyrmion whose 
formation energy is negative. The negative formation energy explains why 
the skyrmion prefers to stretch out to occupy the whole space to lower its energy. 
This process is clearly demonstrated in the movie in the Supplementary Material, 
showing how the system evolves from (a1) to (a3) and how $Q$ grows from 0 to 1. 
The simulations show also that the exact pattern is very sensitive to the initial 
configuration and dynamical/kinetic paths as well as how energy dissipates.  
In a real system, the process shall proceed spontaneously from any randomly generated 
nucleation centre no matter by a thermal fluctuation or an intentional agitation.    

We use color to encode the skyrmion charge density $\rho$ defined as $\rho=\vec m\cdot(\partial_x 
\vec m\times\partial_y \vec m)/(4\pi)$ as indicated by the color bar in Fig. \ref{fig1}. 
Interestingly, $\rho$ is non-zero only around convex (positive) and concave (negative) areas. 
$\rho$ is almost zero along the long straight stripe. This may explain why the skyrmion number 
of stripy textures were thought to be zero in the literature \cite{Wiesendanger}. 
As shown in Fig. \ref{fig1}(a4-c4), the sum of all positive and negative skyrmion charge is 
always quantized to 1 protected by the topology. The continuous decrease of energy $E_{total}$ 
also indicate the final morphology of the ramified stripe skyrmions is not unique and depends 
on dynamical path of the system evolution that in turn relates to the initial configuration.

\begin{figure*}
\centering
\includegraphics[width=17cm]{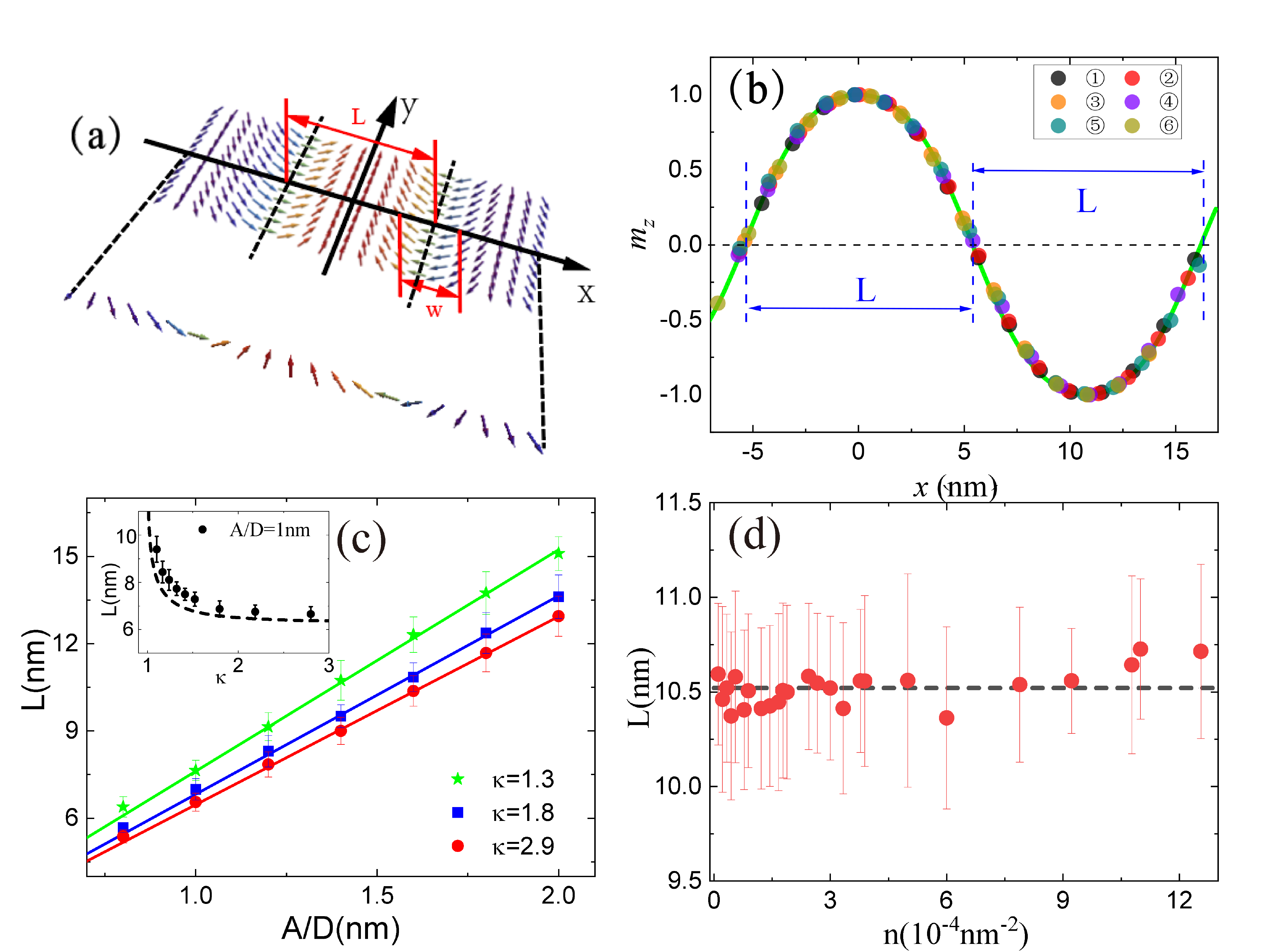}\\
\caption{(a) Schematic diagram of spin texture of parallel stripes. 
(b) Symbols, from different stripes labelled by textcircled{n} 
in Fig. 1(a3-c3), are $m_z=\cos\Theta$ from MuMax3.  
The green solid line 
is the fit of $\Theta (x)=2\arctan\left[\frac{\sinh
(|x|/w)}{\sinh(L/2w)}\right]$. 
($\Theta\leq \pi/2$) and $\Theta (x)=2\arctan\left[\frac{\sinh(L/2w)}{\sinh(
(|x-L|))/w)}\right]$ ($\Theta\geq \pi/2$) with $L=10.78\rm{nm}$ and $w=3.05\rm{nm}$. 
$x=0$ is the centres of white stripes. All data from different stripes fall 
onto the same curve means that stripes of the same width are basic building 
blocks of stripe skyrmions. (c) $A/D$-dependence of stripe width $L$ for various 
$\kappa=\pi^2 D^2/(16AK)=1.3$ (the green stars); 1.8 (the blue squares); and 2.9 (the red circles). 
The solid lines are the fits of $L=aA/D$ with $a=7.61$ for $\kappa=1.3$; $a=6.82$ for $\kappa=1.8$; 
$a=6.47$ for $\kappa=2.9$. Inset: the dependence of $L$ on $\kappa$ for $A/D=1\rm{nm}$. Symbols are the 
numerical data and the dashed line is the theoretical prediction without any fitting parameter. 
(d) The dependences of stripe width $L$ on skyrmion number density for $A=10\rm{pJ/m}, D=6\rm{mJ/m^2}, 
K=0.49\rm{MJ/m^3}$, and $M_s=0.58\rm{MA/m}$. 
The number of skyrmions varies from 1 to 113 in our sample of 
$300\rm{nm}\times 300\rm{nm}\times 0.4\rm{nm}$ under the periodical boundary conditions. 
The dash line is $L=10.52\rm{nm}$.}
\label{fig2}
\end{figure*}

One striking feature about the stripes shown in Fig. \ref{fig1} is a well-defined 
stripe width that is usually referred to a given wave vector in experiments. 
The competition between the exchange energy and DMI energy can easily lead to a $A/D$ 
dependence \cite{Rossler2006, Han2010, Leonov2010, Thiaville2013, Romming2016}. 
However, it is highly non-trivial to understand the effect of the magnetic anisotropy 
on stripe width. In order to understand the underlying physics of this well-defined width,  
one needs a good spin texture profile along 
the direction perpendicular ($x$) to the stripes. 
We found that black ($m_z\leq 0$) and white 
($m_z\geq 0$) stripes can be approximated by $\Theta (x)=2\arctan\left[\frac{\sinh(L/2w)}{\sinh
(|x|/w)}\right]$ and $\Theta (x)=2\arctan\left[\frac{\sinh(|x|/w)}{\sinh(L/2w)}\right]$ ($|x| 
\leq L/2$), respectively. $\Theta$ is the polar angle of the magnetization at position $x$ 
and $x=0$ is the centre of a stripe. $L$ and $w$ measure respectively the stripe width  
and skyrmion wall thickness as schematically illustrated in Fig. \ref{fig2}(a). 
Figure \ref{fig2}(b) demonstrates the excellence of this approximate profile for a 
set of model parameters of $A=10\rm{pJ/m}, D=6\rm{mJ/m^2}, K=0.49\rm{MJ/m^3}$, and $M_s=0.58\rm{MA/m}$. 
The $y-$axis is $m_z$ and $x=0$ is the stripe centre where $m_z=1$. 
Different symbols are numerical data across different stripes labelled by the green 
\textcircled{n} in Fig. \ref{fig1}(a3-c3). The solid curve is the fit of profile $\cos\Theta (x)$ 
with $L=10.78\rm{nm}$ and $w=3.05\rm{nm}$. All data from different stripes falling onto the same curve 
demonstrates that stripes, building blocks of pattern, are identical. 

Using the excellent spin profile, one can obtain magnetic energy of a film filled with the stripe skyrmions 
as a function of $L$ and $w$. Minimizing the energy against $L$ and $w$ allows us to obtain $A$, $D$, and 
$K$ dependence of stripe width $L$ and skyrmion wall thickness $w$. In terms of $L$, $\epsilon=L/(2w)$, 
$\xi=A/D$, and $\kappa$, the total system energy density of a film filled by such stripes is 
\begin{equation}
E=\frac{4D}{\xi}\left[\frac{\xi^2}{L^2}g_1(\epsilon)-\frac{\pi \xi}{4L}+\frac{\pi^2}{64\kappa} 
g_2(\epsilon)\right],
\label{energy-function}
\end{equation}
where $g_1(\epsilon)=\int_0^1 \frac{[2\epsilon \sinh(\epsilon) \cosh(\epsilon x)]^2}{[\sinh^2
(\epsilon)+\cosh^2(\epsilon x)]^2} {\mathrm d}x$, and $g_2(\epsilon)=\int_0^1 
\frac{[2\sinh(\epsilon)\sinh(\epsilon x)]^2}{[\sinh^2(\epsilon)+\cosh^2(\epsilon x)]^2} {\mathrm d}x$.
The optimal stripe width obtained from minimizing energy $E$ is $L=a\frac{4A}{\pi D}$, where $a$ depends 
weakly on $\kappa=\pi^2 D^2/(16AK) >1$. The physics of this result is clear: DMI energy is negative, 
and one can add more stripes by reducing $L$ such that the total energy will be lowered. 
On the other hand, the exchange energy will increase with the decrease of $L$. 
As a result, $L$ is proportional to $A/D$ that is opposite to the behaviour of size of an 
isolated skyrmion whose size increases with $D$ and decreases with $A$ \cite{Xiansi,Wiesendanger}. 
These theoretical results agree very well with micro-magnetic simulations as shown in Fig. 
\ref{fig2}(c) with $a=7.61$ for $\kappa=1.3$; $a=6.82$ for $\kappa=1.8$; $a=6.47$ for $\kappa=2.9$. 
The dependence of $L$ on $\kappa$ for a fixed $A/D=1 \rm{nm}$ is shown in the inset. 
One can see that $L$ depends weakly on large $\kappa\gg 1$. Unexpectedly, $L$ are the same no 
matter whether we have one, two, or more ramified and non-ramified stripe skyrmions. 
This is reflected in the skyrmion number density independence of $L$ as shown in Fig. \ref{fig2}(d).

\begin{figure*}
\centering
\includegraphics[width=17cm]{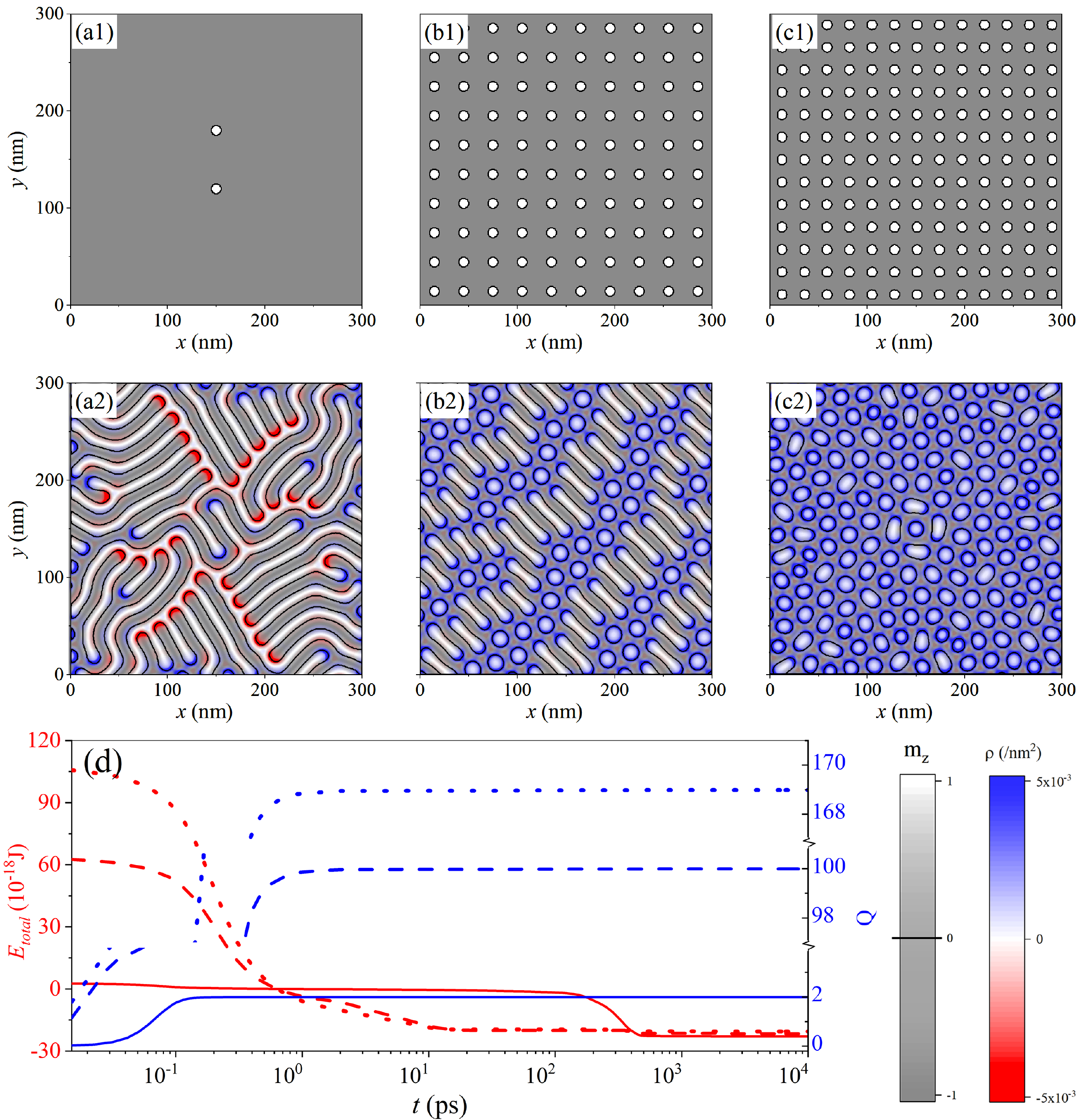}\\
\caption{Evolution of systems with different number of single domains $N=2$ (a), 100 (b) 
and 169 (c) spread in the sample of $300\rm{nm}\times 300\rm{nm}\times 0.4\rm{nm}$ 
with the periodical boundary conditions. The material parameters are the 
same as those in Fig. 1. To speed up the system 
reaching its final stable 
states, we use a large Gilbert damping constant of $\alpha=1$. 
(a1,b1,c1) are the initial configurations and (a2,b2,c2) are the final stable pattern.  
skyrmion charge density $\rho$ is encoded in colours (the red and the blue for 
positive and negative $\rho$ respectively) while $m_z$ is encoded in the grayscale. 
The dark black lines denote $m_z=0$. The positive and negative charges exist respectively 
only around canvex and concave areas. (d) is the evolution of total energy $E_{total}$ 
(the left y-axis) and topological skyrmion number $Q$ (the right y-axis with three ranges). 
Clearly, $Q$ reaches a constant integer in a very short time and $Q=N=2, 100$ and $169$ 
respectively. The total energy is negative and approaches a constant, indicating a stable 
state. }
\label{fig3}
\end{figure*}

To understand why condensed stripe skyrmions and skyrmion crystals can appear 
together in a given chiral magnetic film when the material parameters are fixed, we 
try to increase the number of skyrmions in our film of $300\rm{nm}\times 300\rm{nm}\times 0.4\rm{nm}$.
Encouraged by the results in Fig. \ref{fig1} that each nucleation domain creates one 
irregular stripe skyrmion, we place 2, 100 and 169 small disk domains of diameter $10\rm{nm}$ 
as shown in Fig. \ref{fig3}(a1-c1) and let them to evolve according to LLG equation. 
Fig. \ref{fig3}(a2-c2) are the final steady states. As expected, we indeed obtained 
two irregular ramified stripe skyrmions (a2). For the case of 100 skyrmions, some 
of them have rectangular shape and are arranged in a nematic phase while the rest of 
skyrmions look like disks and are in a lattice structure (b2). In the case of 169 
skyrmions, skyrmions are disk-like and are in a triangular lattice (c2). The skyrmion nature 
of the spin textures can be clearly confirmed by the change of $Q$ (blue) with time as 
shown in Fig. \ref{fig3}(d) (the solid, dashed, and dotted lines for $Q=2,\ 100,$ and 169 
respectively). (d) shows also how system energy (the red lines) changes with time. 
One interesting feature is that total energy is not sensitive to skyrmion number density 
before skyrmion-skyrmion distance is comparable to the optimal stripe width and skyrmions 
take stripe shape. The system energy starts to increase with the skyrmion number density, 
and condensed skyrmions transform from rectangular stripe skyrmions in nematic phase into 
circular skyrmion crystal. This feature does not favour skyrmion crystal formation, and may 
explain why skyrmion crystals were observed in the presence of an external magnetic field: 
With only one skyrmion in the whole system as those shown in Fig. \ref{fig1}, the net 
magnetic moment is zero because spin up (white area) and down spin (black area) are almost 
equal so that magnetic moment cancel each other. As skyrmion number increases, the net 
magnetic moment increases and becomes non-zero. 

We have showed that stripe and ramified stripe skyrmions are essentially the same 
as the circular skyrmions in skyrmion crystals. The difference in skyrmion shapes 
at different skyrmion number density come from the skyrmion-skyrmion interaction. 
When the average distance between two nearby skyrmions is order of the stripe 
width, the skyrmion-skyrmion repulsion compress skyrmions into circular objects. 
This understanding permits a skyrmion crystal in the absence of an external 
magnetic field as long as one can use other means to add more skyrmions 
into a film such as a scanning tunnelling tip \cite{Romming}. \par 

We have presented results for the interfacial DMI so far, similar results are also true for bulk DMI. 
As shown in Fig. 4, one ramified stripe skyrmion  has very similar structures for interfacial (a) and 
bulk (b) DMIs when we start from the same initial configuration with the same interaction strength. The only difference is the change of Neel-type of stripe wall to the Bloch-type stripe wall as shown by the color-coding in the figures. 

A perpendicular magnetic field can modify the morphology and width of a stripe 
without changing its skyrmion number. 
Stripe width increases (decreases) with the 
field  when out-of-plane spin component 
is anti-parallel (parallel) to the field. 

\begin{figure}
\centering
\includegraphics[width=8.5cm]{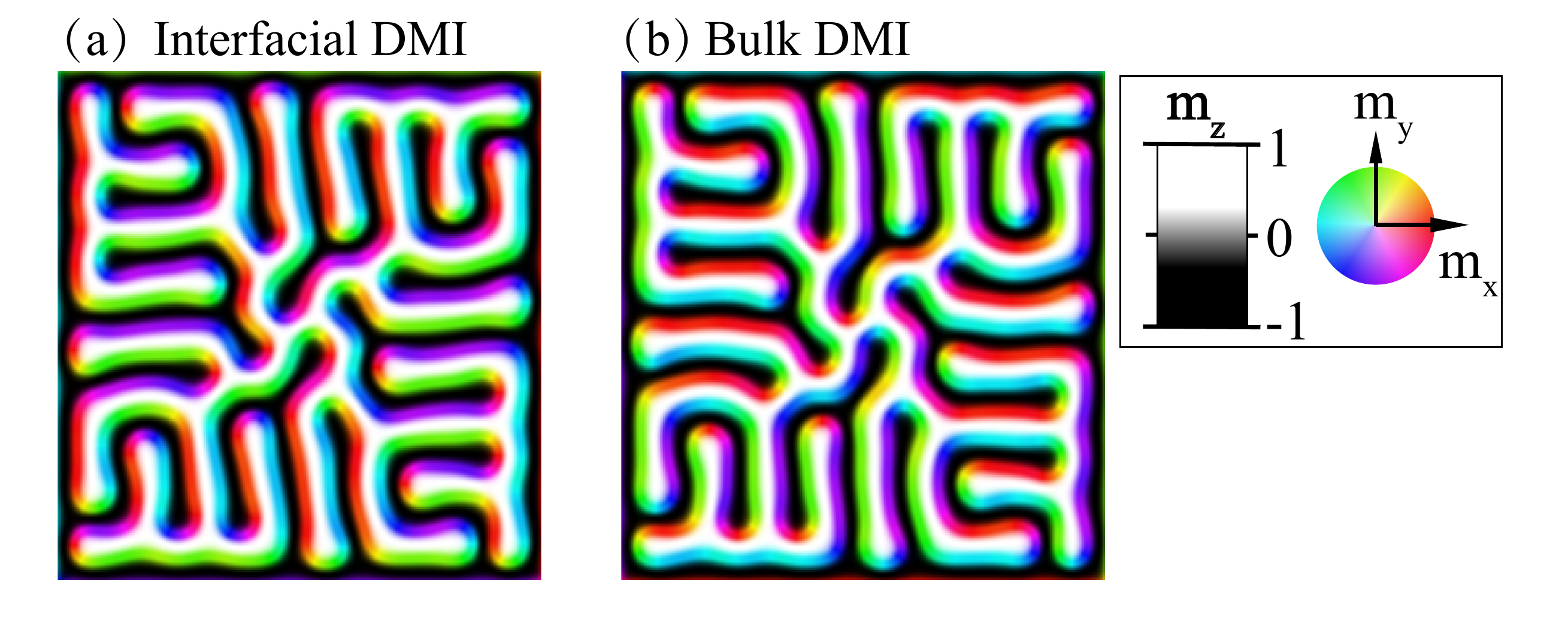}\\
\caption{Structures of one stripe skyrmion 
for interfacial (a) and bulk 
(b) DMIs starting from the same initial configuration. The sample size is 
$300\rm{nm}\times 300\rm{nm}\times 0.4\rm{nm}$ with the same model parameters 
as those for Fig. 1 except $D=4{\rm mJ/m^2}$. The periodical boundary 
conditions are used in the simulations. The in-plane spin component is 
encoded by the colors.}
\label{fig4}
\end{figure}

It requires certain amount energy to destroy a stable or metastable state by definition. Thus, 
all skyrmions discussed in this paper are stable 
against thermal noise as long as the thermal energy is smaller than their potential barriers.  
In the Supplemental Material, we provide a video to 
show that the state in Fig. 3(b2) is stable at 50K. 

Skyrmions provide a fertile ground for studying fundamental physics. 
For example, the topological Hall effect is a phenomenon about how 
non-collinear magnetization in skyrmion crystals affect electron transport. 
Knowing stripy phases are also condensed irregular skyrmions, it expands surely 
arena of topological Hall effect. We can not only study how electron transport 
be affected by the skyrmion crystals, but also skyrmions in other condensed phases 
such as nematic phases, or how the elongation and orientation of stripe skyrmion 
affect Hall transport. With the new discovery of stripe skyrmions, it will also 
allow us to investigate the interplay of topology, shape, spin and charge. 
One can investigate how the topology, the local and global geometries affect 
spin excitations separately.

The assertion that ramified stripes and other stripes appeared together with skyrmion 
crystals are irregular skyrmions are firmly confirmed by the nano-magnetic simulations. 
These stripes have a well-defined width that is from the competition between exchange 
interaction energy and DMI energy. The inverse of the width is the well-known 
wave-vector used to describe the spiral spin order in the literature \cite{Uchida}. 
In contrast to isolated skyrmions whose size increases with DMI constant and decreases  
with the exchange stiffness constant, the stripe width increases with the exchange 
stiffness constant and decreases with the DMI constant. Counter-intuitively, skyrmion 
crystals are highly compressible like a gas, not like an atomic crystal. This detail 
property need a careful further study. We believe that our findings should have 
profound implications in skyrmion-based applications. 

\noindent{\large\textbf{Methods}}\\
\noindent\textbf{Numerical simulations.}
Spin dynamics is governed by the Landau-Lifshitz-Gilbert (LLG) equation,
\begin{equation}
\frac{\partial \vec m}{\partial t} =-\gamma\vec m \times \vec H_{\rm eff} +
\alpha \vec m \times \frac{\partial \vec m}{\partial t},
\label{llg}
\end{equation}
where $\vec m$, $\gamma$, $\alpha$ are respectively the unit vector of the
magnetization, gyromagnetic ratio, and the Gilbert damping.
$\vec H_{\rm eff}=2A \nabla^2\vec m+2Km_z\hat z+\vec H_d+\vec H_{\rm DM}$
is the effective field including the exchange field characterized by
the exchange stiffness $A$, crystalline anisotropy field, demagnetizing field 
$\vec H_d$, and DMI field $\vec H_{\rm DM}$.

In the absence of energy source like an electric current, LLG equation 
describe a dissipative system whose energy can only decrease \cite{xrw1,xrw2}. 
Thus, solving LLG equation is an efficient way to find the stable spin textures. 
In the present case, we apply periodic boundary conditions to eliminate the edge effects.
We use the Mumax3 package \cite{MuMax3} to numerically solve the LLG equation 
with mesh size of $1 \rm{nm}\times 1 \rm{nm}\times 0.4 \rm{nm}$ for those skyrmions of $L> 5 \rm{nm}$.
For skyrmions of $L<5\rm{nm}$ , the mesh size is $0.1 \rm{nm}\times 0.1 \rm{nm} \times 0.4 \rm{nm}$. 
The number of stable states and their structures should not depend on the 
Gilbert damping constant, but spin dynamics is very sensitive to $\alpha$. 
To speed up our simulations, we use large $\alpha$ of 0.25 and 1. 
We consider only material parameters that supports condensed skyrmion states. 
The skyrmion size $L$ is obtained directly from numerical data or by fitting the simulated 
spin profile to $\Theta (x)=2\arctan\left[\frac{\sinh(L/2w)}{\sinh(|x|/w)}\right]$ (the black 
stripes for $m_z\leq 0$) and $\Theta (x)=2\arctan\left[\frac{\sinh(|x|/w)}{\sinh(L/2w)}\right]$ 
(white stripes for $m_z\geq 0$) with stripe width $L$ and wall width $w$.
Here $-L/2\leq x \leq L/2$ and $x=0$ is the centre of a stripe.
It should be pointed out that the physics discussed here does not depend on the boundary conditions. 
However, different boundary conditions have
different confinement potentials (or potential well depth) for stripe skyrmions, and can affect 
the maximal skyrmion number density below which 
a condensed skyrmion state is metastable. 
\\

\noindent\textbf{Energy density of a film filled by stripes.}

Following similar assumptions as those in Ref.~\cite{Thiaville2013}, we can computer the energy density of a 
system filled with stripes of width $L$ that parallel to the $y-$axis and are periodically 
arranged along the $x-$axis. If $\Theta$ is the polar angle of spins and spin profile is
$\Theta(x)=2\arctan\left[\frac{\sinh((|x-nL|)/w)}{\sinh(L/2w)}\right]+n\pi$ with integer 
$n$ and $x\in ((n-0.5)L, (n+0.5)L)$ as shown in Fig. \ref{fig2}(a), the total energy of 
the film in the range of $y_1<y<y_2$ and $x_1<x<x_2$ is 
\begin{equation}
\begin{aligned}
E_{total} =d\int_{y_1}^{y_2}\int_{x_1}^{x_2} [ A(\partial_x\Theta)^2- D \partial_x\Theta+K\sin^2\Theta ] {\rm d}x{\rm d}y. 
\end{aligned}\nonumber
\end{equation}
$x_2-x_1$ and $y_2-y_1$ are much bigger than $L$, we need only to minimize energy density 
$E=E_{total}/[d(x_2-x_1)(y_2-y_1)]$. Then one has 
\begin{equation}
E = \frac{1}{L}\int_{-L/2}^{L/2}\left[ A\left(\frac{\partial\Theta}{\partial x}\right)^2- 
D \frac{\partial\Theta}{\partial x}+K\sin^2\Theta\right] {\rm d}x.
\end{equation}
In terms of $\epsilon=L/(2w)$, terms in $E$ are, 
\begin{equation}
\begin{aligned}
E_{ex} &= \int_{-L/2}^{L/2}A\left[ \frac{\partial\Theta(x)}{\partial x}\right]^2{\mathrm d}x \\
&=\frac{A}{w^2}\int_{-L/2}^{L/2}\left[ \frac{2\sinh(L/w)\cosh(x/w)}{\sinh^2(L/w)+\sinh^2(x/w)}\right] ^2{\mathrm d}x \\
&=\frac{2A}{L}\int_{-1}^{1}\left[ \frac{2\epsilon \sinh(\epsilon)\cosh(\epsilon x)}{\sinh^2(\epsilon)+\sinh^2(\epsilon x)}\right]^2{\mathrm d}x, \\ 
&=\frac{4A}{L}g_1(\epsilon)\nonumber
\end{aligned}
\end{equation}

\begin{equation}
\begin{aligned}
E_{DM} &= - D\int_{-L/2}^{L/2}\frac{\partial\Theta(x)}{\partial x}{\mathrm d}x\\
&=-D\left[\Theta(L/2)-\Theta(-L/2)\right]=-\pi D,
\end{aligned}\nonumber
\end{equation}
\begin{equation}
\begin{aligned}
E_{an} 
&= \int_{-L/2}^{L/2}K \sin^2\Theta (x) {\mathrm d}x\\ &=K\int_{-L/2}^{L/2}\left[ \frac{2\sinh(L/2w)\sinh(x/w)}{\sinh^2(L/2w)+\sinh^2(x/w)}
\right]^2{\mathrm d}x \\ 
&=\frac{K}{2}\int_{-1}^{1}\left[\frac{2 \sinh(\epsilon)\sinh(\epsilon x)}{\sinh^2(\epsilon)+\sinh^2(\epsilon x)}\right]^2L{\mathrm d}x\\ 
&=KLg_2(\epsilon).
\end{aligned}\nonumber
\end{equation}
Add the three terms up, the energy density $E$ is  
$$ E= \frac{4A}{L^2}g_1(\epsilon)-\frac{\pi D}{L}+Kg_2(\epsilon).$$
In terms of $L$, $\epsilon$, $\xi=A/D$ and $\kappa=\pi^2D^2/(16AK)$, we obtain Eq. (2), 
$$E=\frac{4D}{\xi}\left[\frac{\xi^2}{L^2}g_1(\epsilon)-\frac{\pi \xi}{4L}+\frac{\pi^2}{64\kappa} 
g_2(\epsilon)\right].$$
The first and the third terms on the right hand side are positive, thus 
$$E\geq\frac{4D}{\xi}\left[\sqrt{\frac{4\pi^2 \xi^2}{64L^2\kappa}g_1(\epsilon)g_2(\epsilon)}-
\frac{\pi \xi}{4L}\right]=\frac{\pi D}{L}\left[\sqrt{\frac{g_1g_2}{\kappa}}-1\right].$$
To have negative $E$, $\kappa$ must be larger than $\sqrt{g_1g_2}\geq 1$.

\noindent\textbf{Data availability.}
The data that support the plots within this paper and other findings of
this study are available from the corresponding author on reasonable request.
\\

\noindent{\large

\noindent{\large\textbf{Acknowledgement}}\\
This work is supported by Ministry of Science and Technology through grant MOST20SC04-A, the NSFC Grant (No. 11974296 and 11774296)
and Hong Kong RGC Grants (No. 16301518 and 16301619). Partial support by the National Key 
Research and Development Program of China (Grant No. 2018YFB0407600) is also acknowledged.

\noindent{\large\textbf{Author contributions}}\\
X. R. Wang planned the project and wrote the manuscript. 
X.C.H. and H.T.W. performed theoretical and numerical simulations, and prepared the figures. 
All authors discussed the results and commented on the manuscript.

\noindent{\large\textbf{Additional information}}\\

\textbf{Supplementary Information} accompanies this paper
at http://

\noindent{\textbf{Competing interests:}
The authors declare no competing interests.}

\clearpage

\end{document}